# Reference Publication Year Spectroscopy (RPYS) of Eugene Garfield's publications

Lutz Bornmann\*, Robin Haunschild\*\*, Loet Leydesdorff\*\*\*

\* Division for Science and Innovation Studies,

Administrative Headquarters of the Max Planck Society,

Hofgartenstr. 8,

80539 Munich, Germany.

Email: bornmann@gv.mpg.de

\*\* Max Planck Institute for Solid State Research,

Heisenbergstr. 1,

70569 Stuttgart, Germany.

Email: R.Haunschild@fkf.mpg.de

\*\*\* Amsterdam School of Communication Research (ASCoR)

University of Amsterdam,

PO Box 15793,

1001 NG Amsterdam, The Netherlands.

Email: loet@leydesdorff.net

**Abstract**

Which studies, theories, and ideas have influenced Eugene Garfield's scientific work? Recently, the method reference publication year spectroscopy (RPYS) has been introduced, which can be used to answer this and related questions. Since then, several studies have been published dealing with the historical roots of research fields and scientists. The program CRExplorer (http://www.crexplorer.net) was specifically developed for RPYS. In this study, we use this program to investigate the historical roots of Eugene Garfield's oeuvre.

# 1 Introduction

Bibliometrics has become a central component of research evaluation. Field-normalized indicators are used to assess the scientific performance of institutions and countries. Individual researchers are well informed about their h index. The development to this prominence of bibliometrics had not been possible without the groundbreaking work of Eugene Garfield (EG). EG conceptualized the citation indexing in science and published the underlying concept in *Science* (Garfield, 1955). His invention of the index "revolutionized the world of scientific information and made him one of the most visionary figures in information science and scientometrics" (van Raan & Wouters, 2017). In the 1960s, he started the professional use of bibliometrics by founding the Institute for Scientific Information (ISI) (now Clarivate Analytics, see clarivate.com). He developed the Science Citation Index, the Social Sciences Citation Index, and the Arts and Humanities Citation Index. These indexes are predecessors of the Web of Science (WoS) which is well-known today to scholars (Small, 2017; Wouters, 1999b). During EG's lengthy career, he published more than 1500 publications (see www.researcherid.com/rid/A-1009-2008).

In this study, we are interested in the roots of EG's oeuvre. Which studies, theories, and ideas have influenced EG's own scientific work? Marx, Bornmann, Barth, and Leydesdorff (2014) introduced reference publication year spectroscopy (RPYS), which can be used to answer this and related questions. In recent years, several studies have been published dealing with the historical roots of research fields. For example, Hou (2017) deals with the historical roots of citation analysis in a recent study. Thor, Marx, Leydesdorff, and Bornmann (2016a) developed the CRExplorer – a program specifically designed for RPYS (http://www.crexplorer.net). In this study, we use this program to investigate the historical roots of EG's oeuvre.

## 2   Dataset

The study is based on papers published by EG. We used his ResearcherID to collect his publications and added some further publications which are not covered by the ResearcherID. The dataset consists of n=1558 papers published between 1954 and 2016 (containing 15890 cited references, CRs). The papers have been retrieved from the WoS. The downloaded files were imported in the CRExplorer for further processing (Thor, et al., 2016a; Thor, Marx, Leydesdorff, & Bornmann, 2016b). The removing, clustering, and merging functionalities of the CRExplorer have been used to clean the CR dataset, especially for variants of the same CR. The clustering and merging of variants (using volume and page number information) leads to a slightly reduced set of CRs (n=15850). A further 199 CRs were deleted because the bibliographic information was unprecise (for example, no reference publication year, RPY, was provided). We formed two datasets from this initial set for the RPYS:

(1) <u>Dataset including self-citations by EG</u>: We deleted those CRs which have occurrences of less than 10% in one year (i.e. the share of occurrences for a CR among the occurrences of all CRs in one year is less than 10%). This step leads to the first final set of 328 CRs. Note that this also removes two of the most referenced CRs with n=100 and n=97 occurrences as their share of occurrences in their RPYs is 9.9% and 8.3%, respectively. However, they are multiple CRs which appeared in *Current Contents* without a volume number; so they do not provide useful bibliometric information (see below).

(2) <u>Dataset without self-citations</u>: We deleted all self-cited publications (n=1283). Since the CR data in the WoS only contains the first author name, the identification of self-citations is restricted to this first name. As above, we deleted those CRs which have occurrences of less than 10% in a single year (see above). This step results in the second final set of 316 CRs. Although the difference between data set 1 and data set 2 (12 CRs) appears to be rather small at a first glance, the spectrogram changes significantly. Most self-cited CRs (n=1271) are removed due to the occurrence threshold of 10% in a single year.

## 3 Results

The 1558 papers authored by EG are published in 125 different journals, but 68.2% of his papers were published in *Current Contents*. In the 1950s, *Current Contents* has been founded by EG as a rapid alerting service (Small, 2017). The first subject edition covered biology and medicine. Other subject editions were added later, and the original version was renamed into *Current Contents Life Sciences*. Initially, *Current Contents* consisted of printed versions of reproduced title pages of major peer-reviewed journals. An author index with addresses and a subject index were provided as well. This indicates that EG's publications in *Current Contents* have a status different from papers in peer-reviewed journals. Similarly, EG founded *The Scientist* as a professional magazine for scientists in 1986. Although this journal was also incorporated into the *Science Citation Index*, it does not provide references and is thus only registered for being cited; including a Journal Impact Factor (JIF) of 0.369 in 2015.

Table 1 shows the journals in which EG has published at least 10 papers. The journals shown contain 88.6% of EG's publication output.

Table 1. Journals with at least 10 papers published by EG

| Journal | Number of papers |
|---|---|
| *Current Contents* | 1063 |
| *Scientist* | 148 |
| *Current Contents Life Sciences* | 88 |
| *Journal of Information Science* | 13 |
| *Scientometrics* | 12 |
| *Nature* | 12 |
| *Journal of Chemical Documentation* | 12 |
| *Journal of The American Society For Information Science* | 11 |
| *Abstracts of Papers of The American Chemical Society* | 11 |
| *Science* | 10 |

EG has published most of his papers (80.9%) between 1970 and 1990. During these two decades, he has published 60 papers per year on average. However, 77.8% of the papers during this time frame were published in *Current Contents*. The unusual distribution of EG's papers across journals and publication years shows that EG has been an unusual scientist preferring a journalistic medium.

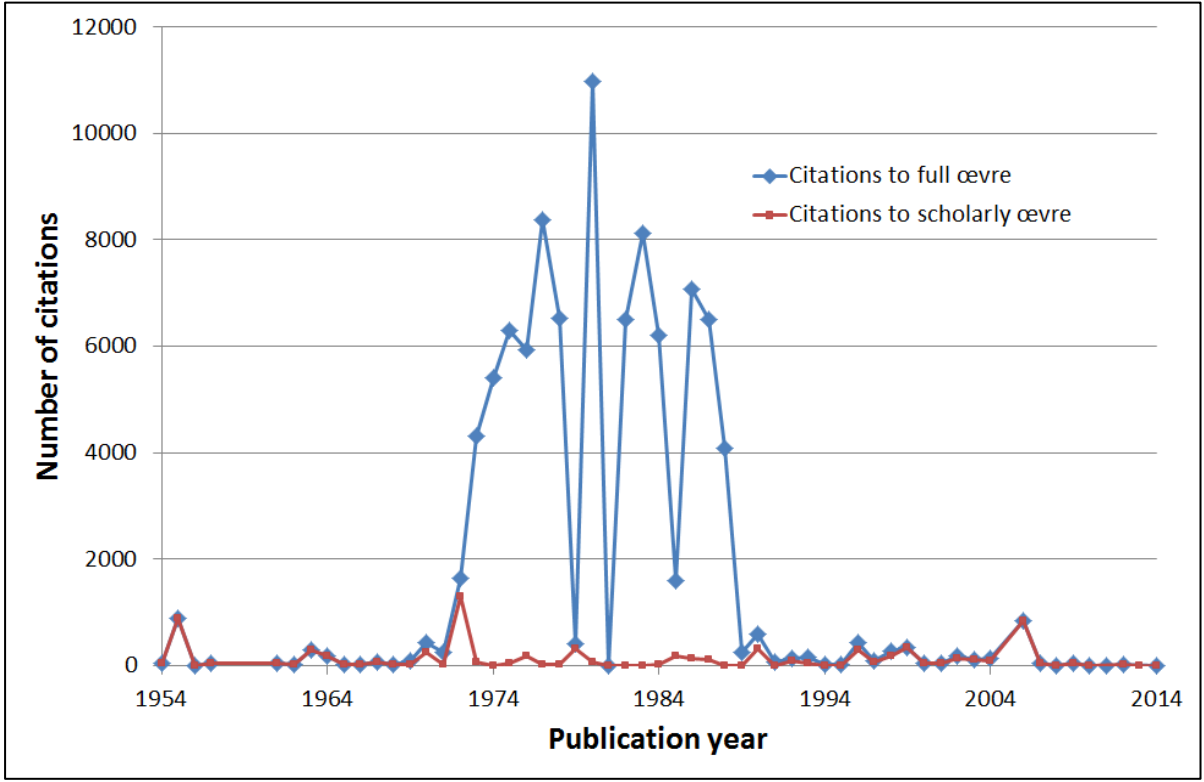

Figure 1: Number of citations to EG's full and scientific oeuvre

Without the publications in *Current Contents* and *The Scientist*, EG's oeuvre is 257 publications in scholarly journals containing 2,244 references. Figure 1 compares the citations to EG's full and scholarly oeuvre. The two papers in *Science* of 1955 and 1972 are heavily cited. The erratic blue-colored curve is mainly citation of *Current Contents* and to a large extent self-citation. The peak in 2006 is an article in *JAMA* entitled "The history and meaning of the journal impact factor".

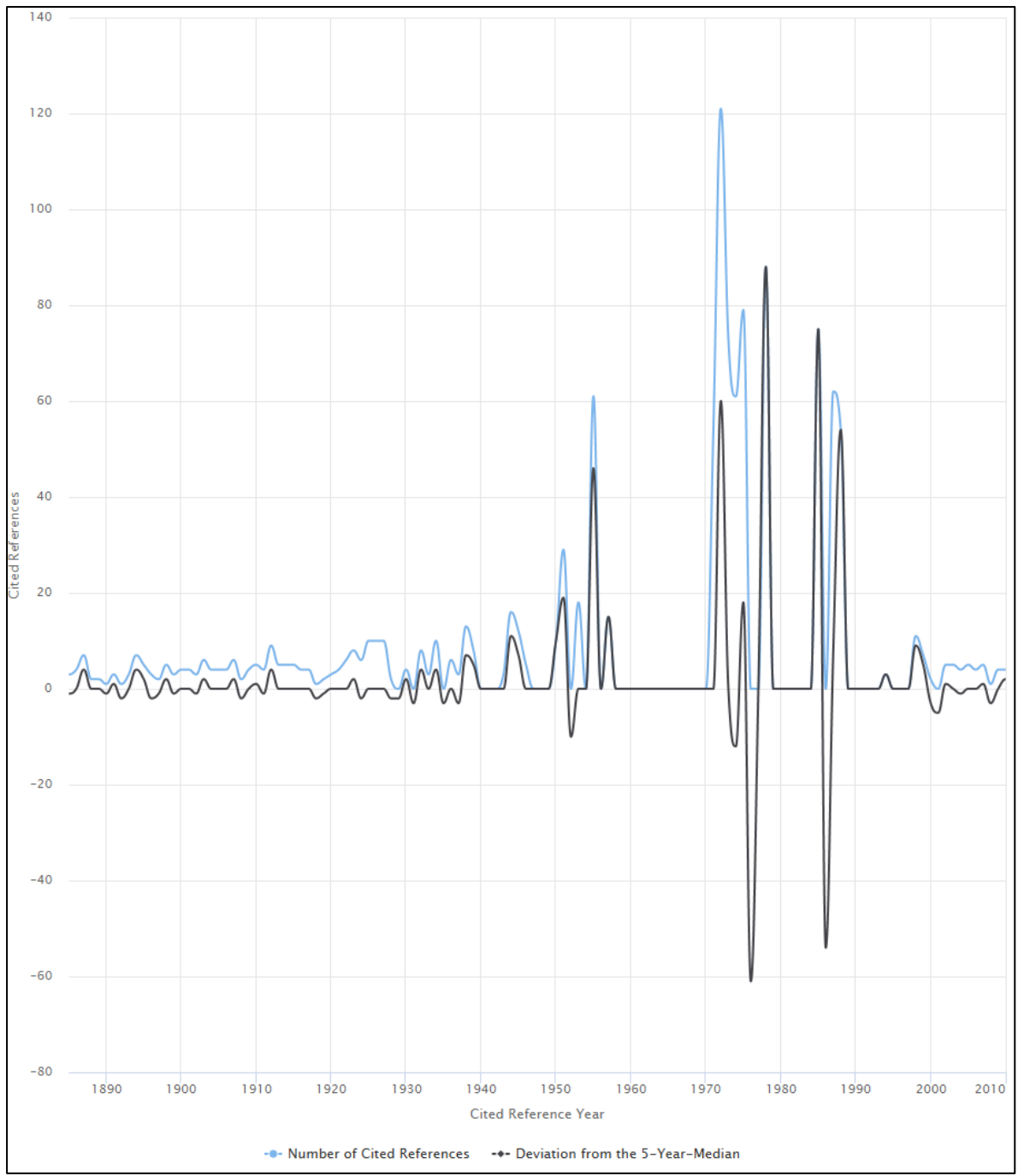

Figure 2. Spectrogram of data set 1 (n=328 CRs). The number of CRs (blue line) and the five-year median deviation (black line) are shown.

The spectrogram of the number of CRs and the five-year median deviations for data set 1 is shown in Figure 2. The spectrogram features five large peaks at RPYs 1955, 1971-1975 (with a shoulder), 1978, 1985, and 1987/1988. Table 3 lists the CRs which are mainly responsible for these five peaks. In addition to the bibliographic information, the table provides the number of occurrences (in other words, how often the publication has been cited by EG) in absolute terms. It is clearly visible that the result is dominated by self-citations. EG seems to have frequently based his newer publications on the footing of his earlier publications. EG was a pioneer and there were no other shoulders than his own, on which he could base his research. Similarly to EG's publication output, also his referencing style is skewed.

Nine of the eleven CRs in Table 2 appeared in *Current Contents*. Unfortunately, no issue or volume numbers are provided in most cases. In 1978, EG has published 52 papers in *Current Contents* with the beginning page number 5—in practice, this is the opening page. Therefore, it is not possible to provide more information about most CRs in Table 2 by bibliometric means; one would have to retrieve and read these contributions. Furthermore, it is unclear which of the volumes/issues was meant by the reference. However, the CR with RPY 1985 (Garfield, E., 1985, *Current Contents*, V43, P3) is about "uses and misuses of citation frequency". The first and the third CR are two papers in *Science.* The 1955 paper introduces the concept of the SCI; the 1972 paper is the *locus classicus* for explaining the JIF. These are the two most cited papers of EG. They both introduce a major innovation.

Table 2. CRs with the largest number of occurrences (including self-citations). The CRs which are responsible for the large peaks in Figure 2 are shown.

| CR | RPY | Number of occurrences |
|---|---|---|
| Garfield, E., 1955, Citation indexes for science - new dimension in documentation through association of ideas. Science, 122(3159), 108-111. | 1955 | 61 |
| Garfield, E., 1971, *Current Contents*, P5 | 1971 | 54 |
| Garfield, E., 1972, Citation analysis as a tool in journal evaluation: journals can be ranked by frequency and impact of citations for science policy studies. Science, 178(4060), 471-479. | 1972 | 64 |
| Garfield, E., 1972, *Current Contents*, 1101, P5 | 1972 | 57 |
| Garfield, E., 1973, *Current Contents*, P5 | 1973 | 73 |
| Garfield, E., 1974, *Current Contents*, P5 | 1974 | 61 |
| Garfield, E., 1975, *Current Contents*, P5 | 1975 | 79 |
| Garfield, E., 1978, *Current Contents*, P5 | 1978 | 88 |
| Garfield, E., 1985, *Current Contents*, V43, P3 | 1985 | 75 |
| Garfield, E., 1987, *Current Contents*, P3 | 1987 | 62 |
| Garfield, E., 1988, *Current Contents*, P3 | 1988 | 54 |

In order to receive hints on publications by other researchers which might have influenced EG's research, we performed a second RPYS after excluding (first author) self-citations (data set 2).

The spectrogram of the number of CRs (blue line) and the five-year median deviations (black line) is shown in Figure 3. The main difference between Figure 2 and Figure 3 is that 12 additional self-cited CRs (after applying the 10% threshold criterion) remain in data set 1 but are removed from data set 2. These 12 additional self-cited CRs are mainly responsible for the shape of the spectrogram from data set 1 (see Figure 2) in comparison to the shape of the spectrogram from data set 2 (see Figure 3). The spectrogram in Figure 3 shows many small peaks and only one distinguished larger peak at the RPY 1951. The reference to Lowry, Rosebrough, Farr, and Randall (1951) is an example used to discuss non-justified extremely high citations (TC>300,000). This is the first CR in Table 3.

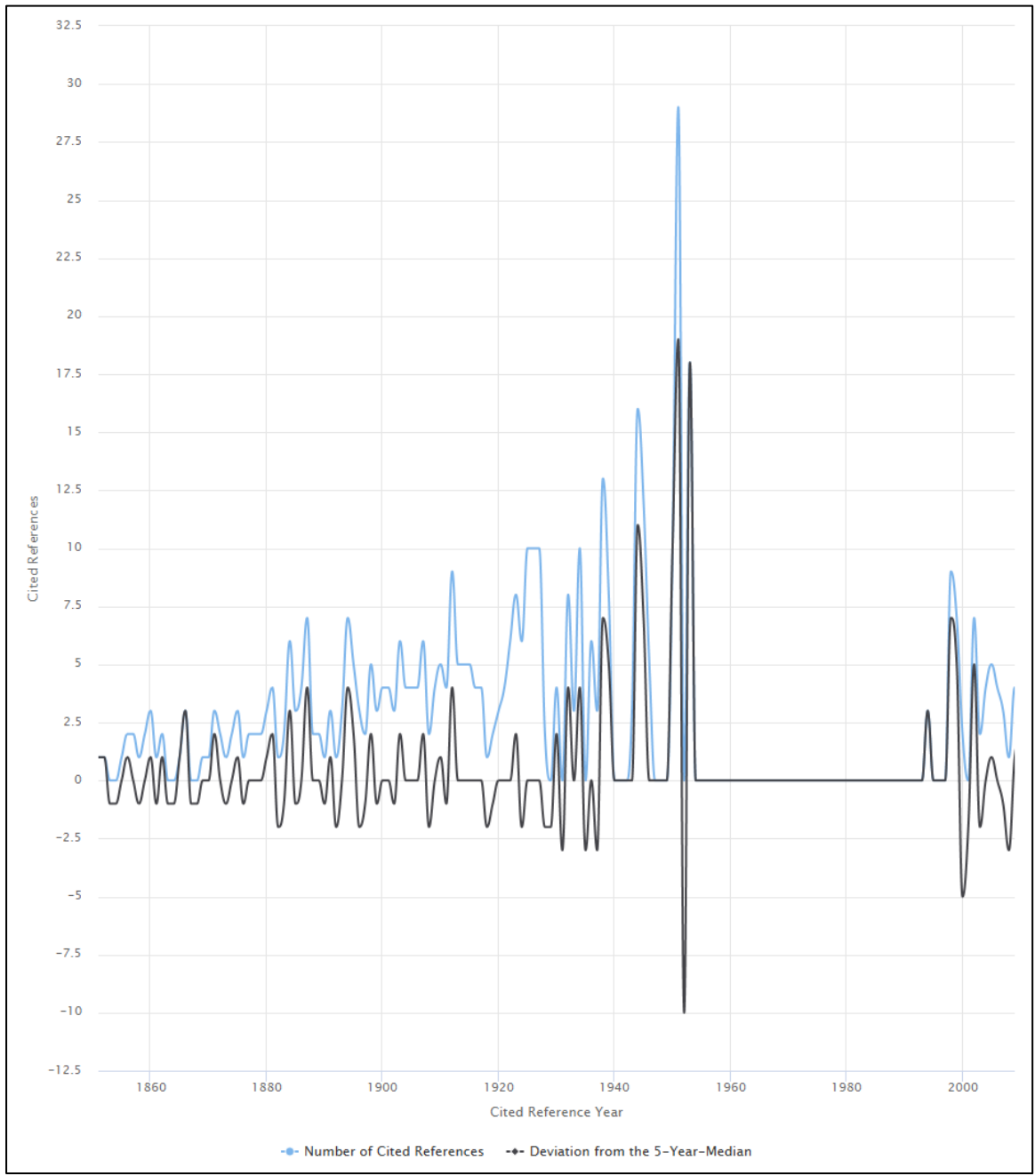

Figure 3. Spectrogram of data set 2 (n=316 CRs). The number of CRs (blue line) and the five-year median deviation (black line) are shown.

We discuss the other important CRs using Table 3 which shows the CRs with the largest number of occurrences within data set 2. On the first glance, we can observe significantly lower numbers of occurrences for the CRs. Thus, these publications were less important for EG to write his papers than his own publications. The CRs in Table 3 can be divided into three groups: (1) bibliometrics, (2) biology, and (3) non-scholarly material.

The group of CRs related to bibliometrics is the largest group in Table 3 with five CRs (the fifth, seventh, eighth, ninth, and tenth CR). The group of CRs which belong to the field of biology is only slightly smaller with four CRs (the first, second, fourth, and eleventh CR). The smallest group of CRs which belongs to non-scholarly material is composed of two CRs (the third and sixth CR). The CRs in Table 3 are ordered by decreasing number of occurrences.

The second and fourth CRs report the discovery (the fourth CR, Avery et al. with RPY=1944) and the structure (the second CR, Watson & Crick with RPY=1953) of Deoxyribose Nucleic Acid (DNA). Watson and Crick described the DNA structure as a double helix in the second CR in Table 3. These authors received the Nobel Prize in 1962 for this work together with Maurice Hugh Frederick Wilkins. The eleventh CR (Selye with RPY=1946) has been referenced by EG three times in *Current Contents*. EG also used this CR as an example paper in his 1955 paper (first CR of Table 3). His 1955 paper was reprinted in 2003 (DOI 10.1093/ije/dyl189) and therefore, counts as two references.

The fifth CR (Bradford, RPY=1950) is foundational for bibliometrics: among other things, it discusses Bradford's (1934) law of scattering. The ninth CR (RPY=1934) is the original publication of Bradford's law of scattering. The seventh CR (RPY=1927) is the proposal by Gross and Gross (1927) to use the number of times a journal is cited as a criterion for purchasing it in the library. Therefore, it can be seen as a predecessor of the JIF. The eighth CR (RPY=1944) is a critical assessment of the aforementioned journal selection criterion using physiology journals as an example. Probably, this CR has fueled EG's pursuit towards the JIF. The tenth CR (RPY=2002) is a paper by Pudovkin and Garfield (2002) where they propose a relatedness factor (RF) for journals calculated from citation counts, number of papers, and number of references. The RF groups journals close together which are also semantically similar.

The third CR (RPY=1945) is an essay about the tasks of scientists after the Second World War. Many aspects of the information society were discussed in this essay. The sixth CR (RPY=1938) is the well-known novel "World Brain" by H. G. Wells.

Table 3. CRs with the largest number of occurrences (without self-citations), ordered by decreasing number of occurrences. CRs with at least 5 occurrences are shown.

| CR | RPY | Number of occurrences |
|---|---|---|
| Lowry, O. H., Rosebrough, N. J., Farr, A. L., & Randall, R. J. (1951). Protein Measurement with the Folin Phenol Reagent. *Journal of Biological Chemistry, 193*(1), 265-275. | 1951 | 29 |
| Watson, J. D., & Crick, F. H. C. (1953). Molecular structure of nucleic acids - a structure for Deoxyribose Nucleic Acid. *Nature*, 171(4356), 737-738. | 1953 | 18 |
| Bush, Vannevar (1945). As we may think. *Atlantic Monthly 176*, 101-108 | 1945 | 12 |
| Avery, Oswald T., MacLeod, Colin M., & McCarty, Maclyn. (1944). Studies on the chemical nature of the substance inducing transformation of pneumococcal types. *Journal of Experimental Medicine, 79*(2), 137-158. | 1944 | 10 |
| Bradford, S. C. (1950). *Documentation*. Washington, DC, USA: Public Affairs Press. | 1950 | 10 |
| Wells, H.G. (1938). *World Brain*. London. UK: Methuen. | 1938 | 9 |
| Gross, P. L. K., & Gross, E. M. (1927). College libraries and chemical education. *Science, 66*, 385-389. | 1927 | 8 |
| Brodman, Estelle. (1944). Choosing Physiology Journals. *Bulletin of the Medical Library Association, 32*(4), 479-483. | 1944 | 6 |
| Bradford, S. (1934). On the scattering of papers on scientific subjects in scientific periodicals. Engineering, 137(193), 85. | 1934 | 6 |
| Pudovkin, A. I., & Garfield, E. (2002). Algorithmic procedure for finding semantically related journals. *Journal of the American Society for Information Science and Technology, 53*(13), 1113-1119. | 2002 | 5 |
| Selye, H. (1946). The general adaptation syndrome and the diseases of adaptation. *Journal of Clinical Endocrinology & Metabolism, 6*, 117-231. | 1946 | 5 |

## 4    Discussion

Beyond doubt, EG has been one of the most influential scientists in the field of bibliometrics. The field would perhaps not exist, if EG had not introduced the idea of indexing citations for quantitative analyses in the history and philosophy of science (Elkana, Lederberg, Merton, Thackray, & Zuckerman, 1978; Gingras, 2016; Wouters, 2000). EG created the fundament on which all empirical studies in bibliometrics are based. In addition to the idea of citation indexing, he initiated several lines of research in bibliometrics later on. For example, the development of indicators based on bibliometric data started with the introduction of the JIF (Garfield, 2006; Garfield & Sher, 1963). Since this initial introduction, the development of several (advanced) indicators followed in bibliometrics which do not only refer to journals (several journal-based indicators can be found in Clarivate Analytics' Journal Citation Report), but also to other entities (e.g., to single researchers with the h index).

EG was the first who formulated a list of possible reasons why documents are cited in publications (Bornmann & Daniel, 2008). Thus, he started the discussion in bibliometrics about the meaning of citations and the formulation of underlying theoretical approaches. A further example of his groundbreaking work is the introduction and development of algorithmic historiography (Garfield, 2004; Garfield, Sher, & Torpie, 1964; Leydesdorff, 2010). The method which can be applied by using the historiographic software HistCite™ "facilitates the understanding of paradigms by enabling the scholar to identify the significant works on a given topic" (Garfield, Pudovkin, & Istomin, 2003, p. 400). Similar software which has been developed on this base later on are CitNetExplorer (http://www.citnetexplorer.nl, see van Eck & Waltman, 2014) and CRExplorer (www.crexplorer.net, see Thor, et al., 2016a, 2016b).

This study deals with EG's roots (studies, theories, and ideas) and how they are reflected in EG's publications. We used the CRExplorer – which has its roots in EG's own software developments – and applied RPYS to his publication set. The results show that EG has cited his own publications in many cases. This can be possibly traced back to the fact that EG was a pioneer not only in shaping the citation index, but also in many lines of research in bibliometrics. Furthermore, inspiring inputs might have been come from non-citable sources. For example, Shepard's Citations is a citation index which is used in US legal research. It lists all authorities who cite particular cases, statutes, or other legal authorities. EG later acknowledged that his citation index was heavily influenced by Shepard's Citations (Wouters, 1999a). However, this does not reflect in (or cannot be retrieved from) his citations, because there does not exist a paper on Shepard's Citations which could be cited in this context.

Among the papers most referenced by EG are some non-scholarly publications, some publications from the early evolving field of bibliometrics, and publications from other disciplines for the presentation of examples. The heterogeneity of the cited publications demonstrates the wide-spreading interests of EG which range from the historiography of science as an empirical operationalization for questions in the history and philosophy of science (e.g. Elkana, et al., 1978; Gingras, 2016), citation indexing as an approach to document retrieval (Garfield, 1955), journal evaluation for library management (Bradford, 1950; Garfield, 1957, 1972), the mapping of the sciences (Garfield, 1987; Small & Garfield, 1985), to bibliometrics and the theory of citation (Garfield, 1975, 1979).

The prevailing publication and citation of non-scholarly publications makes clear that EG's interests are not limited by scholarly bounds. They reach beyond the narrow definitions of scientific epistemological interests and perform the translation of the latter into knowledge-based innovations.